\documentclass[12pt]{iopart}

\usepackage{iopams}  
\usepackage{graphicx}
\usepackage{color}
\usepackage{cite}
\usepackage[squaren]{SIunits}

\newcommand{\dif}{\mathrm{d}}%
\renewcommand{\vec}[1]{\boldsymbol{#1}}%
\newcommand{\CC}{\boldsymbol{\mathcal{C}}}%
\newcommand{\DD}{\boldsymbol{\mathcal{D}}}%
\newcommand{\HH}{\boldsymbol{\mathcal{H}}}%
\newcommand{\MM}{\boldsymbol{\mathcal{M}}}%
\newcommand{\uu}{\boldsymbol{\hat{u}}}%
\newcommand{\ee}{\boldsymbol{\hat{e}}}%

\begin{document}

\title[Can anisotropic microswimmers be described by using forces and torques?]{Can the self-propulsion of anisotropic microswimmers be described by using  forces and torques?}

\author{Borge ten Hagen$^{1,*}$, Raphael Wittkowski$^2$, Daisuke Takagi$^3$, Felix K\"ummel$^4$, Clemens Bechinger$^{4,5}$ and Hartmut L\"owen$^1$}
\address{$^1$ Institut f\"ur Theoretische Physik II: Weiche Materie, Heinrich-Heine-Universit\"at D\"usseldorf, D-40225 D{\"u}sseldorf, Germany}
\address{$^2$ SUPA, School of Physics and Astronomy, University of Edinburgh, Edinburgh EH9 3FD, United Kingdom}
\address{$^3$ Department of Mathematics, University of Hawaii at Manoa, Honolulu, Hawaii
96822, USA}
\address{$^4$ 2.\ Physikalisches Institut, Universit{\"a}t Stuttgart, D-70569 Stuttgart, Germany}
\address{$^5$ Max-Planck-Institut f{\"u}r Intelligente Systeme, D-70569 Stuttgart, Germany}
\vskip1ex\address{$^*$ E-mail: bhagen@thphy.uni-duesseldorf.de}

\date{\today}

\begin{abstract} 
The self-propulsion of artificial and biological microswimmers (i.e.,
active colloidal particles) has often been modelled by using 
a force and a torque entering into the overdamped equations for the Brownian motion of passive particles.
This seemingly contradicts the fact that a swimmer is force-free and torque-free, i.e., that the net force and torque on the particle vanish. Using different models for mechanical and diffusiophoretic
self-propulsion, we demonstrate here that the equations of motion of microswimmers can be mapped onto those
of passive particles with the shape-dependent grand resistance matrix and formally external \textit{effective} forces and torques. This is consistent with experimental findings on the
circular motion of artificial asymmetric microswimmers driven by self-diffusiophoresis. 
The concept of effective self-propulsion forces and torques significantly
facilitates the understanding of the swimming paths, e.g., for a microswimmer under gravity.
However, this concept has its limitations when the self-propulsion mechanism of a swimmer is disturbed either by another particle in its close vicinity or by interactions with obstacles, such as a wall. 
\end{abstract}


\bibliographystyle{unsrt}

\section{\label{Einleitung}Introduction}
The basic interest in liquid matter research has shifted from passive 
simple liquids \cite{hansen-mcdonald:86} to complex liquids \cite{Barrat_book} over the past decades.
While the individual building blocks of the liquid are passive thermal particles 
in this case, a more recent topic concerns \textit{living liquids} which consist of self-propelling 
biological constituents, such as bacteria in a low-Reynolds-number fluid \cite{catesreview,marchettireview,Wensink_PNAS,Poon}. 
Being perpetually driven by the consumption of energy needed for the self-propulsion, even a single 
swimmer represents already a complicated non-equilibrium situation.
In this respect, artificial self-phoretically-driven colloidal particles serve as 
very helpful model systems to mimic the self-propulsion of biological microswimmers 
and enable particle-resolved insights into the one-particle and collective phenomena
\cite{Palacci2010,VolpeSM1,theurkauff2012dynamic,BechingerJPCM,Speck_PRL2013,Kuemmel:13,Palacci2013,tenHagenKWTLB2014}.

Regarding the modelling of microswimmer propulsion, there are basically two levels of description.
The first coarse-grained approach does not resolve the details of the propagation but
models the resulting propulsion velocity $\vec{v}$ by the action of a formally external \textit{effective} self-propulsion force 
$\vec{F}$ moving with the particle. This force then enters into the completely overdamped equations of motion for passive particles. In a demonstrative way, $\vec{F}$ can be interpreted as the constraining force that is required to prevent the microswimmer from moving, which can be achieved, e.g., by optical tweezers \cite{Brady2014}. 
In its simplest form, i.e., for a sphere of hydrodynamic radius $R$, the self-propulsion force $\vec{F}$ is given by 
\begin{equation}%
\gamma \vec{v}  = \vec{F}  
\label{eq:v0}%
\end{equation}\noindent%
with $\gamma= 6\pi \eta R >0$ denoting 
a (scalar) Stokes friction coefficient for stick fluid boundary conditions on the surface of the particle and
$\eta$ denoting the dynamic (shear) viscosity of the fluid. 
There is plenty of literature by now in which this kind of force modelling was 
employed either in the context of single-particle motion \cite{Teeffelen_PRE,tenHagenWL2011,WittkowskiL2012,Elgeti:13} or to study various collective phenomena of self-propelled particles such as swarming \cite{Chen2006,Li2008,Wensink2012}, clustering \cite{Peruani2006,Mehandia2008,Wensink:08,McCandlish2012,Bialke_PRL2012,Palacci2013,Baskaran_PRL2013,Fily2014}, and ratchet effects \cite{Angelani2011,Reichhardt:13}. The concept has also been used for several applications like particle separation \cite{Yang2012,Costanzo2014} or trapping \cite{KaiserPRL:12,Kaiser:13,Wang2014} of self-propelled particles.
Typically, the swimming direction is along a particle-fixed axis (usually a symmetry axis of the particle) denoted by a unit vector
$\uu$. The direction of the self-propulsion force thus rotates with the particle orientation. Therefore it is called an `internal' force. The particle orientation $\uu$ obeys an overdamped equation of motion with rotational friction.
This makes the problem non-Hamiltonian and non-trivial, in particular in 
the presence of noise \cite{Howse:07,tenHagen_JPCM}. However, throughout this article thermal fluctuations are neglected as they are not relevant for the present study.

From (\ref{eq:v0}) it is evident how to generalize this equation towards an asymmetric microswimmer with an arbitrary non-spherical shape by using 
the $6\!\times\!6$-dimensional grand resistance matrix $\HH$ \cite{HappelB1991} (also called `hydrodynamic friction tensor' \cite{Kraft:13}) and a six-dimensional generalized velocity $\vec{V}=({\vec v}, {\vec \omega})$
composed of the translational velocity $\vec{v}$ and the angular velocity $\vec{\omega}$ of the particle.
Correspondingly, the right-hand side of (\ref{eq:v0}) is replaced by the six-dimensional generalized force $\vec{K}=(\vec{F},\vec{M})$ composed of an internal force $\vec{F}$ and an internal torque $\vec{M}$, which are fixed in the particle frame, such that \footnote{In two spatial dimensions the grand resistance matrix $\HH$ is only $3\!\times\!3$-dimensional and the generalized velocity $\vec{V}$ as well as the generalized force $\vec{K}$ are three-dimensional.}
\begin{equation}
\eta \HH \vec{V} = \vec{K}.
\label{eq:v0gen}
\end{equation}
Equation (\ref{eq:v0gen}) comprises many different situations including circle swimming in two \cite{Teeffelen_PRE,Kuemmel:13} and helical swimming in three dimensions \cite{WittkowskiL2012}, which are induced by the simultaneous presence of a constant force and a torque in the particle frame. Many papers have adopted this formalism in order to describe the motion of microswimmers 
\cite{Teeffelen:09,Cebers2011,Grossmann2012,Radtke2012,Si2012,Ganguly2013,Llopis2013,Marine2013,Takagi2013,Crespi2013,KaiserPRE2013,Volpe2013,Martinelli2014,Mallory2014,Babel2014,Brady2014,StenhammarMAC2014}.
On this coarse-grained level of modelling there is no explicit solvent velocity field, which is just disregarded.

The second level involves a more detailed description of the propulsion mechanism of an individual swimmer.
At this level, several modes of propagation have to be distinguished. We shall discriminate in the following between
mechanical self-propulsion and self-diffusiophoretic motion.
Some type of mechanical self-propulsion is usually realized in biological microswimmers such as bacteria \cite{Frymier1995,DiLuzio2005,Sokolov2010}, algae \cite{Polin2009,Kantsler2013}, or spermatozoa \cite{Riedel2005,Friedrich2008,Elgeti2010,Friedrich2010JEB}. In particular, flagellar locomotion \cite{Lighthill1976,Purcell1997,Lauga_review:09} has been studied intensely. But also different propulsion mechanisms have been analysed theoretically \cite{Swan2011,Lobaskin2008}. In general, the active motion is induced by mutually moving objects such as rotating screws \cite{Wada2006} or non-reciprocally translated spheres
as embodied in the paradigmatic three-sphere swimmer of Najafi and Golestanian \cite{Najafi2004}. 
Diffusiophoretic self-propulsion is typically described by imposing a slip solvent velocity on the particle surface \cite{Anderson1989,Ajdari2006},
which then puts the swimmer into motion \cite{Paxton,Downton:09,Goetze:10,Pagonabarraga2010}. A similar way of modelling is also well-established for squirmers \cite{Lighthill1952}, where a tangential surface velocity is prescribed \cite{Ishikawa2006,Ishikawa2008,Alarcon2013,MatasNavarro2014}.
In both cases, the swimming mode itself is intrinsic and therefore occurs in the absence of any external force or torque,  
which is typically expressed by the fact that a swimmer is force-free and torque-free \cite{Purcell:77}. 
Clearly, on this level of description the solvent velocity field enters explicitly.

At first sight, the description on the first level using internal forces and torques seems to be in contradiction to 
the fact that a swimmer is force-free and torque-free. It has therefore been 
criticized \cite{Felderhof}. 
However, in fact there is no such contradiction if the 
self-propulsion force and torque in the equations of motion are interpreted as formal or effective quantities \cite{KuemmeltHWBEVLB2014}. For spherical particles, the magnitude of the effective force just fixes the propulsion speed via the formal relation (\ref{eq:v0}). We emphasize that this concept is not capable of revealing any insights about the detailed self-propulsion mechanism itself. In particular, it may not be confused with the disputable assumption of an osmotic driving force \cite{CordovaFigueroa2008,JuelicherProstPRL} on a microscopic level. How the effective force has to be interpreted in the context of a mechanical self-propulsion mechanism was made explicit in
\cite{GolestanianEPJE} for the motion of a three-sphere swimmer with one big and two small spheres on a linear array.
The force contribution acting on the big sphere due to two point forces located at the positions of the small spheres is not given as a simple sum of these two forces. Instead, they have to be appropriately rescaled first, in order to provide the effective force which obeys (\ref{eq:v0}).

The issue becomes less evident for anisotropic swimmers which perform in general a circular (in two dimensions) or helical (in three dimensions) motion and whose dynamics is governed by a translational-rotational coupling \cite{WittkowskiL2012,Kuemmel:13}. 
The basic question covered in this article is whether a similar formal rescaling can be done
in order to map the equations of motion of an asymmetric microswimmer onto those of a corresponding passive particle described by (\ref{eq:v0gen}).
Or in other words: can the grand resistance matrix $\HH $ with formal forces and torques $\vec{K}$ be used to 
describe the self-propulsion of microswimmers? From linear response theory it is clear that the leading part of the
equations is linear such that there is a matrix relating the generalized velocities 
$\vec{V}$ to the generalized forces $\vec{K}$. 
But is this matrix identical to the grand resistance matrix of the corresponding passive particles? This has recently been put into question by Felderhof
\cite{Felderhof}. This important issue can be answered either by experiments or by a detailed theoretical analysis.
In \cite{Kuemmel:13} it was shown that the experimental data for the planar motion of asymmetric microswimmers can indeed be described using
the grand resistance matrix occurring in (\ref{eq:v0gen}).

In this article we follow the alternative theoretical route: 
we model the swimmer motion explicitly according to the more detailed 
second level  and check whether the resulting equation
can be mapped onto the coarse-grained equation (\ref{eq:v0gen}) using an effective force and torque. We do this explicitly
for an asymmetric two-dimensional swimmer with an L-shape and discriminate between mechanical self-propulsion and 
self-diffusiophoretic propulsion in an unbounded fluid. In the first case, we consider an extension of the Golestanian swimmer \cite{Najafi2004,GolestanianEPJE} towards an L-shaped particle. In the second case, we study slender biaxial particles consisting of two arms of different lengths that meet at a certain internal angle. To account for the diffusiophoretic self-propulsion mechanism, we consider an imposed slip velocity on one of the arms. The sense of rotation of the particles is found to depend on the internal angle. 
A previous but less general account for a rectangular internal angle was 
published elsewhere \cite{KuemmeltHWBEVLB2014,tenHagenKWTLB2014}.
For both the mechanical and the diffusiophoretic self-propulsion, we explicitly show that the matrix needed to describe the motion of the microswimmer is identical to the grand resistance matrix of the particle. This opens the way to think in terms of 
effective forces and torques, which is advantageous in order to get insights into the variety of swimming paths in more complicated situations. The benefit becomes most obvious when both external body forces and torques and internal effective self-propulsion forces and torques are present, such as in gravitaxis \cite{tenHagenKWTLB2014,Wolff2013,campbell2013gravitaxis}.

The paper is organized as follows: the mechanical self-propulsion mechanism and a generalization of the three-sphere swimmer of Golestanian towards an L-shaped microswimmer are discussed in section \ref{mechanisch}. Afterwards, in section \ref{diffusiophoretisch}, the self-propulsion of biaxial diffusiophoretic microswimmers is studied using slender body theory. Based on these considerations for two classes of microswimmers, section \ref{general} contains general considerations for asymmetric particles with arbitrary shape and arbitrary self-propulsion. In section \ref{limitations}, we discuss limitations of the concept of effective forces and torques before we finally conclude in section \ref{Schluss}.

\section{\label{mechanisch}Mechanical self-propulsion}
Microswimmers with mechanical self-propulsion change their shapes in a non-reciprocal way \cite{Purcell:77} in order to propel themselves forward. 
These shape changes can be realized by deformations or -- if the microswimmer consists of several individual parts -- by translating or rotating different parts relative to each other through internal forces and torques $\vec{k}^{(\alpha)}$ that are applied on the individual parts of the microswimmer. 
The noise-free equations of motion of such a mechanical microswimmer with $n$ parts can be formulated through a modified force and torque balance (see, e.g., (42) and (45) in \cite{Higdon}) 
\begin{equation}
\vec{K}_{\mathrm{St}}+\vec{K}=\vec{0} \,,\qquad \vec{K}=\sum^{n}_{\alpha=1}\CC^{(\alpha)}\vec{k}^{(\alpha)}
\label{eq:FM_balance_mech}%
\end{equation}
that ensures that the microswimmer is force-free and torque-free and that includes the Stokes friction force and torque $\vec{K}_{\mathrm{St}}=-\eta\HH\vec{V}$, which is proportional to the shape-dependent \footnote{Note that the grand resistance matrix depends only on the shape but not on the activity of the particle. It is therefore the same for active and passive particles as long as the activity is not accompanied by a change of the shape.} grand resistance matrix $\HH$ of the particle, and the \textit{rescaled} internal forces and torques $\CC^{(\alpha)}\vec{k}^{(\alpha)}$ with rescaling matrices $\CC^{(\alpha)}$. 
For a specific particle shape, the rescaling matrices $\CC^{(\alpha)}$, which take into account the hydrodynamic interactions between the individual parts of the microswimmer, can in principle be calculated using the Green's function approach \cite{Higdon,Kim_Karrila}.
It is important to note that due to this rescaling the equations of motion of the microswimmer are given through a \textit{modified} and not through a simple direct force and torque balance.
The sum of the rescaled internal forces and torques is the formally external \textit{effective} force and torque vector $\vec{K}$.

The simplest mechanical microswimmer is the linear three-sphere swimmer of Najafi and Golestanian \cite{Najafi2004}, which consists of three spheres one behind the other that are translated relative to each other (see (5) in \cite{GolestanianEPJE} for the equation of motion).  
In the following, we generalize their theoretical approach to an L-shaped microswimmer in two spatial dimensions in order to provide a minimal microscopic model for an asymmetric mechanical microswimmer. 

\begin{figure}[tbh]
\centering
\includegraphics[width=0.5\columnwidth]{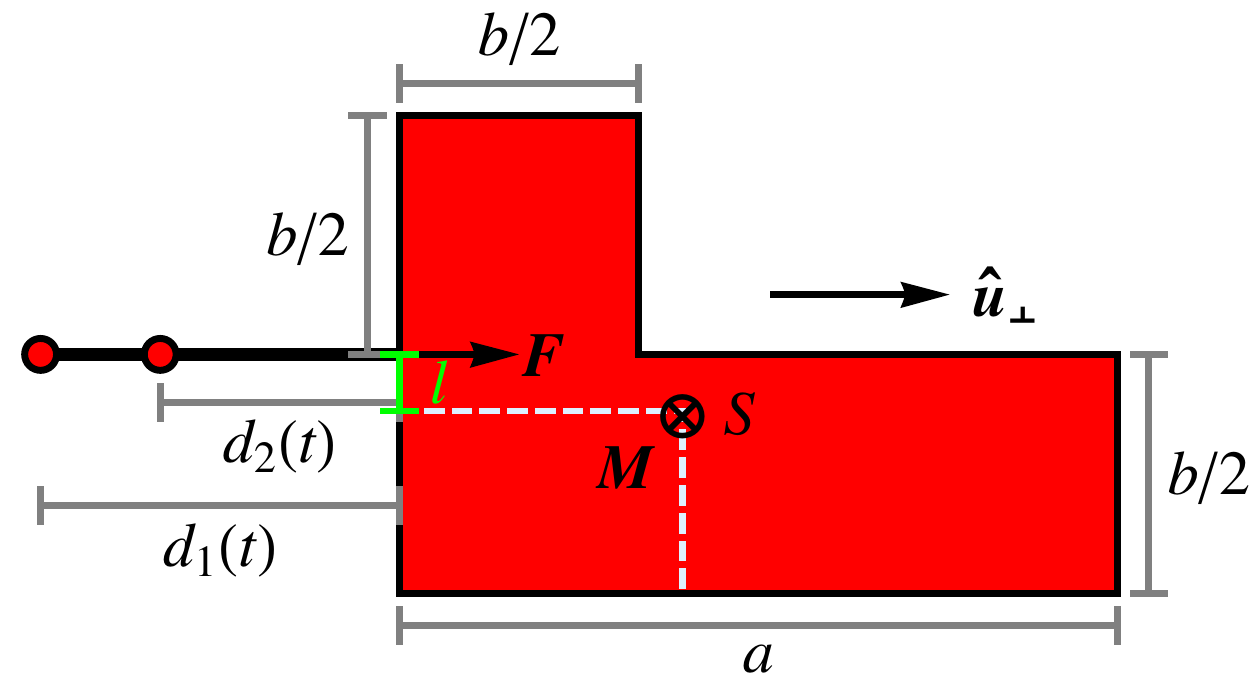}
\caption{\label{figmech}Schematic of a force-free and torque-free L-shaped Golestanian-like microswimmer of size $a$ and orientation $\uu_{\perp}$. To put the microswimmer in motion, the distances $d_{1}(t)$ and $d_{2}(t)$ of the two small spheres from the L-shaped particle vary as a non-reciprocal function of time. The effective lever arm $l$ relates the effective force $\vec{F}$ to the effective torque $M=||\boldsymbol{M}||$ on the centre of mass $S$.
}
\end{figure}

Our L-shaped Golestanian-like microswimmer consists of two very small spheres and a much larger L-shaped particle (see figure \ref{figmech}). 
The small spheres have distances $d_{1}(t)$ and $d_{2}(t)$ from the short arm of the L-shaped particle, which vary as a non-reciprocal function of time (see \cite{Najafi2004} for details) as a consequence of internal forces $\vec{f}^{(\alpha)}$ with $\alpha\in\{1,2\}$ that act on the two spheres. Note that the forces $\vec{f}^{(\alpha)}$ are two-dimensional here as the motion is restricted to the $x$-$y$ plane.
We denote the length of the long arm of the L-shaped particle by $a$ and its orientation by the unit vector $\uu_{\perp}$. 
The length of the short arm of the particle is $b$ and the width of the arms is $b/2$.  
The small spheres are only translated relative to the larger L-shaped particle. There are no internal torques that rotate the spheres and the L-shaped particle relative to each other. 
Nevertheless, the resulting effective (here three-dimensional) generalized self-propulsion force $\vec{K}=(\vec{F},M)=\CC^{(1)}\vec{f}^{(1)}+\CC^{(2)}\vec{f}^{(2)}$ includes not only an effective self-propulsion force $\vec{F}$, but also an effective self-propulsion torque $M$ that acts on the centre of mass $S$ of the particle. 
Since the effective force $\vec{F}$ acts perpendicularly on the middle of the short arm of the L-shaped particle, the effective torque is given by $\vec{M}=lF\ee_z$ with $\ee_z=(0,0,1)$ and $F=||\vec{F}||$. The effective self-propulsion torque $\vec{M}$ can thus be expressed by the effective lever arm $l$, which denotes the distance in the direction of the short arm of the particle between the point on which the effective force acts and the point of reference for the calculation of the diffusion coefficients, which is the centre of mass in our case.
 
In two spatial dimensions, the grand resistance matrix of the L-shaped microswimmer, which in the limit of very small spheres is equal to the grand resistance matrix of the (passive) L-shaped part of the microswimmer, is $3\times 3$-dimensional; the generalized velocity is the three-dimensional vector $\boldsymbol{V}=(\dot{x},\dot{y},\dot{\phi})$ with $\phi$ indicating the particle orientation, and the internal forces $\vec{f}^{(\alpha)}$ are two-dimensional (there are no internal torques). Lastly, the rescaling matrices $\CC^{(\alpha)}$ are $3\times 2$-dimensional. 
With this notation, the force and torque balance (\ref{eq:FM_balance_mech}), which comprises the conditions of zero external force and zero external torque, reduces to 
\begin{equation}
\eta \HH  \boldsymbol{V} = \sum\limits_{\alpha=1}^2 \CC^{(\alpha)} \vec{f}^{(\alpha)} 
\label{eq:force-free}%
\end{equation}
with the generalized effective force 
\begin{equation}
\vec{K}=\Bigg(\!\!\!\begin{array}{c}\vec{F}\\M\\\end{array}\!\!\!\Bigg)=\sum\limits^2_{\alpha =1} \CC^{(\alpha)} \vec{f}^{(\alpha)} \,.
\label{eq:defK}
\end{equation}
From (\ref{eq:force-free}) it is obvious that the equations of motion for the mechanically driven microswimmer studied here can be written in terms of the grand resistance matrix $\HH$ and effective forces and torques as defined in (\ref{eq:defK}). The specific shape of the particle enters into the equations of motion via the grand resistance matrix $\HH$ and the rescaling matrices $\CC^{(\alpha)}$. In principle, these quantities can be calculated for arbitrary particle shapes. However, in the context of experiments with active particles, the knowledge of the exact relations for the effective force and torque is usually not required as these quantities can directly be obtained from the measured trajectories \cite{tenHagenKWTLB2014}.

\section{\label{diffusiophoretisch}Diffusiophoretic self-propulsion}
\subsection{\label{slender}Slender rigid particles}
In contrast to mechanical microswimmers, self-diffusiophoretic microswimmers do not change their shapes in order to propel themselves forward. 
They instead move as a consequence of a local concentration gradient, which they generate themselves. 
The self-propulsion of such self-diffusiophoretic microswimmers can be modelled by prescribing a non-vanishing slip velocity on (parts of) the surface of the particle \cite{Anderson1989,Ajdari2006}. Interestingly, such a hydrodynamic modelling leads to equations of motion that are on a more coarse-grained model the same as the equations of motion of the corresponding passive particles complemented by effective forces and torques to model the self-propulsion.    
In the following, we prove this for biaxial particles with two arms of different lengths, which meet at a certain internal angle, through a hydrodynamic calculation based on slender-body theory for Stokes flow \cite{Batchelor,CoxSBT}.

\begin{figure}[tbh]
\centering
\includegraphics[width=0.4\columnwidth]{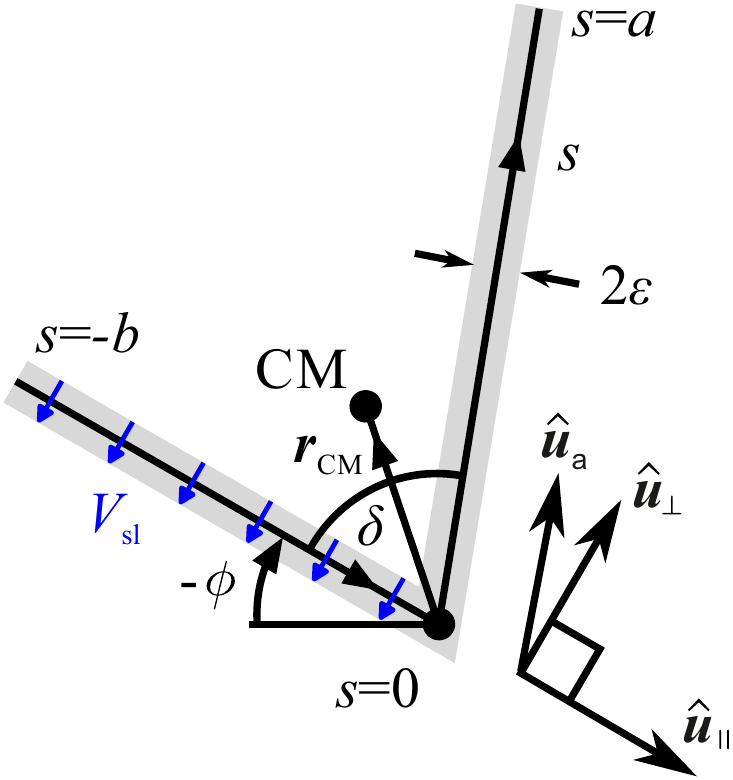}
\caption{\label{figdiff}Sketch of a rigid biaxial particle with two straight arms of lengths $a$ and $b$, the internal angle $\delta$, the centre of mass CM, and a prescribed fluid slip velocity $V_\mathrm{sl}$ on the arm of length $b$. 
The particle-fixed rectangular coordinate system is indicated by the unit vectors $\uu_\parallel=(\cos(\phi),\sin(\phi))$ and $\uu_\perp=(-\sin(\phi),\cos(\phi))$, where $\uu_\parallel$ is parallel to the arm of length $b$, whereas the orientation of the arm of length $a$ is denoted by the unit vector $\uu_{\mathrm{a}}=(-\cos(\delta-\phi),\sin(\delta-\phi))$.}
\end{figure}

Applications of slender-body theory include Purcell's three-link swimmer \cite{Purcell:77,Becker2003} as well as the modelling of flagellar locomotion in general \cite{Lighthill1976,Lauga_review:09}. 
In this section we apply slender-body theory to a rigid slender particle with two arms of different lengths as sketched in figure \ref{figdiff}. 
The two straight arms of lengths $a$ and $b$ enclose the internal angle $\delta$ with $0<\delta<\pi$. The special case $\delta=\pi/2$ corresponds to an L-shaped particle similar to what has been studied, using a different approach, in section \ref{mechanisch}. The results of the slender-body approach for this special case have already briefly been presented in \cite{KuemmeltHWBEVLB2014}. 
Here, we provide a more general and detailed derivation, which shows that the concept of effective forces and torques follows intrinsically from a self-propulsion mechanism that involves a prescribed slip velocity. As required by slender-body theory, the sum $L=a+b$ of the lengths of the two arms of the particle is assumed to be significantly larger than the width $2\epsilon$ of the arms. Although this is not always the case in related experiments, the physical structure of the equations of motion governing the dynamics of asymmetric microswimmers is clearly elucidated by the slenderness approximation $\epsilon\ll L$.

In order to describe the translational and rotational motion of the biaxial microswimmer, we define the centreline position  
\begin{eqnarray}
\vec{x}(s)= \vec{r}-\vec{r}_{\mathrm{CM}} +s\vec{x}' \qquad\text{for}\; -b\le s\le a \,.
\label{centreline}%
\end{eqnarray}
Here, $s$ is the arc length coordinate, $\vec{r}$ is the particle's centre-of-mass position, the vector $\vec{r}_{\mathrm{CM}}=(a^2\uu_{\mathrm{a}}-b^2\uu _\parallel)/(2L)$ is directed from the meeting point of the arms to the centre of mass CM (see figure \ref{figdiff}), and $\vec{x}'=\partial\vec{x}/\partial s$ is the unit tangent, which is $\uu_\parallel=(\cos(\phi),\sin(\phi))$ along the arm of length $b$ ($-b\le s\le 0$) and $\uu_{\mathrm{a}}=(-\cos(\delta-\phi),\sin(\delta-\phi))$ along the arm of length $a$ ($0\le s \le a$). 
For slender arms with width $2\epsilon\ll L$, the fluid velocity at any point on the particle surface can be approximated by $\dot{\vec{x}}(s)+\vec{v}_{\mathrm{sl}}(s)$, where $\dot{\vec{x}}(s)=\partial\vec{x}/\partial t$ is the time derivative of $\vec{x}$ and $\vec{v}_{\mathrm{sl}}$ is the local slip velocity averaged over the perimeter of the surface adjacent to $\vec{x}(s)$. 
We consider a constant slip velocity $\vec{v}_{\mathrm{sl}}=-V_{\mathrm{sl}}\uu_\perp$ pointing in the direction $-\uu_\perp$ normal to the centreline along the arm of length $b$ and a vanishing slip velocity along the other arm. The fluid velocity on the particle surface 
is related to the local force per unit length $\vec{f}(s)$ by the leading-order slender-body approximation
\begin{equation}
\dot{\vec{x}}+\vec{v}_{\mathrm{sl}}=c \, (\boldsymbol{\mathcal{I}}+\vec{x}'\!\otimes\!\:\!\vec{x}')\vec{f}\,,
\end{equation}
where $c=\log(L/\epsilon)/(4\pi \eta)$ is a constant depending on the viscosity $\eta$ of the solvent, $\boldsymbol{\mathcal{I}}$ is the identity matrix, and $\otimes$ denotes the dyadic product. This approximation is valid as long as $\delta$ is not too small so that lubrication effects in the corner between the two arms are negligible. 
The total force on the particle is given by
\begin{equation}
\vec{F}_\mathrm{ext}=\int_{-b}^a \!\!\!\!\vec{f}\,\dif s
\label{force2}%
\end{equation}
and the total torque on the particle is
\begin{equation}
M_\mathrm{ext}=\ee_{z}\!\cdot\!\!\!\int_{-b}^a \!\!\!\! \left(-\tilde{\vec{r}}_{\mathrm{CM}} +s\tilde{\vec{x}}'\right) \!\times\!\tilde{\vec{f}}\,\dif s \,.
\label{torque2}%
\end{equation}
To define the vector product $\times$ in (\ref{torque2}), we technically add a third (zero) component to the vectors $\vec{r}_{\mathrm{CM}}$, $\vec{x}'$, and $\vec{f}$, which is indicated by the tilde.
For force-free and torque-free microswimmers, $\vec{F}_\mathrm{ext}$ and $M_\mathrm{ext}$ are zero, whereas they are non-zero for externally driven particles.

The integral constraints (\ref{force2}) and (\ref{torque2}) lead to a system of three dynamic equations for $\uu _\parallel\!\cdot\!\vec{r}$, $\uu_{\mathrm{a}}\!\cdot\!\vec{r}$, and $\phi$. 
We begin by differentiating (\ref{centreline}) with respect to time and multiplying with $\uu_\parallel$ and $\uu_{\mathrm{a}}$, respectively. This leads to 
\begin{equation}
\fl \uu _\parallel\!\cdot\!\dot{\vec{r}}+\frac{a^2\sin(\delta)}{2L}\dot{\phi}=\cases{2c\:\!\uu _\parallel\!\cdot\!\vec{f}\,, & if $-b\le s\le 0$\,, \\
c\:\!(\uu _\parallel\!\cdot\!\vec{f}-\cos(\delta)\uu_{\mathrm{a}}\!\cdot\!\vec{f})+s\sin(\delta)\dot{\phi}\,, & if $0< s\le a$ }
\label{n}%
\end{equation}
and 
\begin{equation}
\fl \uu_{\mathrm{a}}\!\cdot\!\dot{\vec{r}}+\frac{b^2\sin(\delta)}{2L}\dot{\phi}= \cases{c\:\!(\uu_{\mathrm{a}}\!\cdot\!\vec{f}-\cos(\delta)\uu _\parallel\!\cdot\!\vec{f})-s\sin(\delta)\dot{\phi}+V_\mathrm{sl}\sin(\delta)\,, \!\!\!\!\!\!\!\!\!\!\! & if $-b\le s\le 0$\,, \\
2c\:\!\uu_{\mathrm{a}}\!\cdot\!\vec{f}\,, \!\!\!\!\!\!\!\!\!\!\! & if $0< s\le a$\,.\\}
\label{m}%
\end{equation}
In order to eliminate the unknown $\vec{f}$ from (\ref{n}) and (\ref{m}), the equations are integrated over $s$, separately from $-b$ to $0$ and from $0$ to $a$. After that, (\ref{force2}) and (\ref{torque2}) are used to complete the system of differential equations. To apply the torque condition (\ref{torque2}), equations (\ref{n}) and (\ref{m}) are first multiplied by $s$ before integrating over the two sections. Thus, one obtains eight equations altogether. Taking suitable linear combinations of these equations finally leads to the dynamic equations
\begin{equation}
\eta\HH\left(\begin{array}{c}
\uu _\parallel\!\cdot\!\mathbf{\dot{r}} \\
\uu_{\mathrm{a}}\!\cdot\!\mathbf{\dot{r}} \\
\dot{\phi}
\end{array}\right) 
= \left(\begin{array}{c}
\uu _\parallel\!\cdot\!\vec{F}_\mathrm{ext} \\
\uu_{\mathrm{a}}\!\cdot\!\vec{F}_\mathrm{ext} \\
M_\mathrm{ext}
\end{array}\right)+
V_\mathrm{sl} \left(\begin{array}{c}
0 \\
b\sin(\delta)/c \\
ab(a\cos(\delta)-b)/(2cL)
\end{array}\right)
\label{gov}%
\end{equation}
with the grand resistance matrix
\begin{equation}
\HH=\frac{1}{2 c \eta}\left(\begin{array}{ccc}
2a+b & a\cos(\delta) & \frac{ab(b\cos(\delta)-a)\sin(\delta)}{2L} \\
b\cos(\delta) & a+2b & \frac{ab(a\cos(\delta)-b)\sin(\delta)}{2L} \\
\frac{-a^2 b\sin(\delta)}{2L} & \frac{-ab^2\sin(\delta)}{2L} & A
\end{array}\right)
\end{equation}
and the constant
\begin{equation}
A=(a^3+b^3)\left(\frac{2}{3}-\frac{ab \sin^2(\delta)}{4L^2}\right)-\frac{a^4+b^4+2a^2b^2\cos(\delta)}{2L} \,.
\end{equation}
These equations of motion describe the microswimmer in the non-rectangular particle's frame of reference defined by the unit vectors 
$\uu _\parallel$ and $\uu_{\mathrm{a}}$ and can be interpreted as follows. For force-free and torque-free biaxial swimmers driven by a fluid slip velocity $V_\mathrm{sl}$ along one of the arms (see figure \ref{figdiff}) the first vector on the right-hand side of (\ref{gov}) vanishes. For passive particles driven by an external force $\vec{F}_\mathrm{ext}$ and torque $M_\mathrm{ext}$ the second vector on the right-hand side of (\ref{gov}) vanishes. With regard to the equations of motion the two cases are equivalent if $\uu _\parallel\!\cdot\!\vec{F}_\mathrm{ext} = 0$, $\uu_{\mathrm{a}}\!\cdot\!\vec{F}_\mathrm{ext}=V_\mathrm{sl}b\sin(\delta)/c$, and $M_\mathrm{ext}=V_\mathrm{sl}ab(a\cos(\delta)-b)/(2cL)$, although the respective hydrodynamic flow fields are clearly different \cite{Ramachandran2006,JuelicherEPJE}. This means that the motion of the self-propelled particle with slip velocity $V_\mathrm{sl}$ can be described using an effective force and an effective torque in combination with the same grand resistance matrix $\HH$ as for a corresponding passive particle.

As an additional result, we remark that the sense of rotation of the force-free and torque-free self-propelled particle switches when $b=a\cos(\delta)$. For a specific particle shape with the critical angle $\delta_\mathrm{crit}=\arccos (b/a)$, the particle translates without rotation. For $0<\delta<\delta_\mathrm{crit}$, i.e., if the orthogonal projection of the arm with length $a$ onto the arm with length $b$ is longer than the latter arm, the particle rotates counter-clockwise, whereas for $\delta_\mathrm{crit}<\delta<\pi$ the particle rotates clockwise.

We now consider the special case of an L-shaped particle, i.e., $\delta=\pi/2$. In this special case 
the unit vectors $\uu_\perp$ and $\uu_{\mathrm{a}}$ are equal and the grand resistance matrix $\HH$ in (\ref{gov}) becomes  \cite{KuemmeltHWBEVLB2014}
\begin{eqnarray}
\HH_\mathrm{L} = \frac{1}{2 c \eta}
\left(\begin{array}{ccc}
2a+b & 0 & -a^2b/(2 L)\\
0 & a+2b & -ab^2/(2 L) \\
-a^2 b/(2 L) & -ab^2/(2 L) & A_\mathrm{L} 
\end{array}\right)
\end{eqnarray} 
with the constant
\begin{equation}
A_\mathrm{L}=(a^3+b^3)\left(\frac{2}{3}-\frac{ab}{4L^2}\right)-\frac{a^4+b^4}{2L} \,.
\end{equation} 
Thus, the equations of motion of a force-free and torque-free L-shaped diffusiophoretic microswimmer read
\begin{equation}
\eta \HH_\mathrm{L}
\left(\!\begin{array}{c}
\uu_\parallel\!\cdot\!\mathbf{\dot{r}} \\
\uu_\perp\!\cdot\!\mathbf{\dot{r}} \\
\dot{\phi}
\end{array}\!\right) 
= V_\mathrm{sl} \left(\!\begin{array}{c}
0 \\
b/c \\
-ab^2/(2cL)
\end{array}\!\right) 
\label{govL}%
\end{equation}
in the particle frame. The grand resistance matrix $\HH_\mathrm{L}$ in (\ref{govL}) can be expressed in terms of the generalized diffusion tensor 
\begin{equation}
\DD_0=\frac{1}{\beta\eta}\HH_\mathrm{L}^{-1} = \left(\begin{array}{ccc}
D_{\parallel} &  D_{\parallel}^{\perp} & D^{\parallel}_{\mathrm{C}} \\
D_{\parallel}^{\perp} & D_{\perp} & D^{\perp}_{\mathrm{C}} \\
D^{\parallel}_{\mathrm{C}} & D^{\perp}_{\mathrm{C}} & D_{\mathrm{R}}
\end{array}\right). 
\end{equation} 
In order to transform the equations of motion (\ref{govL}) from the particle's frame of reference to the laboratory frame, we use the rotation matrix 
\begin{equation}
\boldsymbol{\MM}(\phi)= \left(\begin{array}{ccc}
\cos \phi & - \sin \phi & 0 \\
\sin \phi &  \cos \phi & 0 \\
0 & 0 & 1 
\end{array}\right). 
\end{equation}
With this definition, the generalized diffusion tensor $\DD(\phi)$ in the laboratory frame of reference is given by
\begin{equation}
\DD(\phi)=\MM(\phi) \DD_0 \MM^{-1}(\phi)\,.
\end{equation}
Using the orientation vectors $\uu_\parallel$ and $\uu_\perp$, it can also be written as
\begin{equation}
\DD(\phi)= \left(\begin{array}{cc}
\DD_\mathrm{T}(\phi) & \vec{D}_\mathrm{C}(\phi) \\
\vec{D}^\intercal_\mathrm{C}(\phi) & D_{\mathrm{R}}
\end{array}\right)
\end{equation}  
with the translational short-time diffusion tensor
$\DD_{\mathrm{T}}(\phi)=\,D_{\parallel}\uu_{\parallel}\otimes\uu_{\parallel}
+D^{\perp}_{\parallel}(\uu_{\parallel}\otimes\uu_{\perp}+\uu_{\perp}\!\otimes\uu_{\parallel})
+D_{\perp}\uu_{\perp}\!\otimes\uu_{\perp}$, the translational-rotational coupling vector
$\vec{D}_{\mathrm{C}}(\phi)=D^{\parallel}_{\mathrm{C}}\uu_{\parallel}+D^{\perp}_{\mathrm{C}}\uu_{\perp}$ and its transpose $\vec{D}^\intercal_\mathrm{C}(\phi)$, and the rotational diffusion coefficient $D_{\mathrm{R}}$.
Thus, one finally obtains the equations of motion for an L-shaped self-propelled particle
\begin{eqnarray}%
\dot{\vec{r}}&= \beta F \big[\DD_{\mathrm{T}}(\phi) \uu _{\perp}+l\vec{D}_{\mathrm{C}}(\phi) \big] \;, \nonumber \\
\dot{\phi}&= \beta F\big[l D_{\mathrm{R}} +\vec{D}_{\mathrm{C}}(\phi)\!\cdot\!\uu _{\perp}\big]  \,,
\label{eq:LangevinGLG}%
\end{eqnarray}%
where $F=b V_\mathrm{sl}/c$ is the effective self-propulsion force and $l=-ab/(2L)$ is an effective lever arm \cite{Kuemmel:13}.

The above hydrodynamic calculation shows that also for a self-diffusiophoretic microswimmer the concept of effective forces and torques can be applied so that its equations of motion are obtained directly and much more easily from the equations of motion of the corresponding 
passive particle by introducing effective forces and torques that model the self-propulsion.

\subsection{\label{grigid}General rigid particles}
On top of the previous derivation for an L-shaped particle based on slender body theory, we now describe how the effective forces and torques can in principle be calculated for an arbitrarily shaped particle propelled by self-diffusiophoresis. The procedure is based on the Lorentz reciprocal theorem for low-Reynolds-number hydrodynamics \cite{HappelB1991}. It was similarly applied in \cite{Stone1996} in order to relate the translational and rotational velocities of a swimming microorganism to its surface distortions \cite{GonzalezRodriguez2009,Lauga2014}. Such surface deformations and a streaming of the cell surface have been suggested as the swimming mechanism of cyanobacteria \cite{Waterbury1985}, for example. The modelling is similar to diffusiophoresis, where usually a slip velocity at the particle surface is assumed in the theoretical description \cite{Anderson1989,Ajdari2006}. 

In order to calculate the effective forces and torques for an arbitrarily shaped rigid particle with a surface $S$ and vectors $\vec{s}$ that define the points on $S$, we first study how the slip velocity $\vec{v}_\mathrm{sl}(\vec{s})$ on the surface affects the translational and rotational velocities $\vec{v}$ and $\vec{\omega}$ of the particle. On the one hand, we consider the velocity field $\vec{u}$ and the stress field $\vec{\sigma}$ around a rigid self-propelled particle that translates at velocity $\vec{v}$ due to some slip velocity $\vec{v}_\mathrm{sl,t}(\vec{s})$ on the particle surface $S$. On the other hand, we assume that $\tilde{\vec{u}}$  and $\tilde{\vec{\sigma}}$ are the velocity and stress fields around a corresponding passive particle  that moves with velocity $\tilde{\vec{v}}$ due to an external force $\vec{F}_\mathrm{ext}$. The Lorentz reciprocal theorem states that 
\begin{equation}
\int_S \vec{n}\cdot\tilde{\vec{\sigma}}\cdot\vec{u}\,\dif S=\int_S \vec{n}\cdot\vec{\sigma}\cdot\tilde{\vec{u}}\,\dif S\,,
\label{eq:Lorentz}
\end{equation}
where $\vec{n}$ is the unit outward normal to the surface $S$ and $\vec{n}\cdot\vec{\sigma}$ is the stress exerted by the fluid on the surface. 
Note that the right-hand side of (\ref{eq:Lorentz}) vanishes because $\tilde{\vec{u}}=\tilde{\vec{v}}$ is constant on the surface of the passive particle in the second case and $\int_S  \vec{n}\cdot\vec{\sigma}\,\dif S=\vec{0}$ because the self-propelled particle in the first case is force-free. Setting $\vec{u}=\vec{v}+\vec{v}_\mathrm{sl,t}$ on the surface of the self-propelled particle, we obtain
\begin{equation}
\label{force}
\int_S \vec{n}\cdot\tilde{\vec{\sigma}}\cdot(\vec{v}+\vec{v}_\mathrm{sl,t})\,\dif S=0\,.
\end{equation}
This indicates that $\vec{v}$ is directly related to $\vec{v}_\mathrm{sl,t}$. What is required is the stress $\vec{n}\cdot\tilde{\vec{\sigma}}$ on the passive particle, which can in principle be calculated by solving the Stokes equations for the flow around the particle.  Equation (\ref{force}) can also be written as
\begin{equation}%
\vec{F}_\mathrm{ext}\cdot\vec{v}+\int_S \vec{n}\cdot\tilde{\vec{\sigma}}\cdot\vec{v}_\mathrm{sl,t}\,\dif S=0\,.%
\label{eq:Fv}%
\end{equation}%
Using a similar application of the Lorentz reciprocal theorem for a rigid self-propelled particle that rotates at angular velocity $\vec{\omega}$ due to some slip velocity $\vec{v}_\mathrm{sl,r}(\vec{s})$ on the particle surface $S$ and a corresponding passive particle that rotates with angular velocity $\tilde{\vec{\omega}}$ due to an external torque $\vec{M}_\mathrm{ext}$, one can derive the relation 
\begin{equation}%
\vec{M}_\mathrm{ext}\cdot\vec{\omega}+\int_S \vec{n}\cdot\hat{\vec{\sigma}}\cdot\vec{v}_\mathrm{sl,r}\,\dif S=0\,,
\label{eq:Momega}%
\end{equation}%
where $\vec{n}\cdot\hat{\vec{\sigma}}$ is the stress exerted by the fluid on the particle surface. 
By decomposing the slip velocity $\vec{v}_\mathrm{sl}$ into the parts that translate and rotate the self-propelled particle, $\vec{v}_\mathrm{sl}=\vec{v}_\mathrm{sl,t}+\vec{v}_\mathrm{sl,r}$, one could in principle predict the velocities $\vec{v}$ and $\vec{\omega}$. Although (\ref{eq:Fv}) is only a one-component equation, all components of $\vec{v}$ are accessible because different realizations of the external force $\vec{F}_\mathrm{ext}$ and the corresponding stress fields $\tilde{\vec{\sigma}}$ can be inserted. Analogously, $\omega$ can be obtained from (\ref{eq:Momega}). 
Once the translational and rotational velocities are known, the effective force $\vec{F}$ can be obtained by integrating the stress $\tilde{\vec{\sigma}}\cdot\vec{n}$ for the special case $\tilde{\vec{v}}=\vec{v}$ over the particle surface $S$. Correspondingly, the effective torque $\vec{M}$ is the integral of $(\vec{s}-\vec{r}_{0}) \times (\hat{\vec{\sigma}}\cdot\vec{n})$ for the special case $\tilde{\vec{\omega}}=\vec{\omega}$ over the particle surface $S$, where $\vec{r}_{0}$ is the reference point for the torque (e.g., the centre of mass of the particle). Following this procedure, the effective forces and torques for an arbitrarily shaped self-propelled particle can be calculated in principle.

\section{\label{general}Considerations for general self-propulsion}
Based on the previous explicit investigations of active particles with either mechanical or diffusiophoretic self-propulsion, here we briefly present some general considerations about effective forces and torques on a rigid self-propelled particle. For these general considerations, the particle shape and the origin of the self-propulsion do not matter.

Assume that a rigid self-propelled particle with surface $S$ moves with translational velocity $\vec{v}$ and angular velocity $\vec{\omega}$. The only relevant condition is that the net force and the net torque vanish as required for swimmers \cite{Purcell:77}. On the other hand, we consider a passive rigid particle with the same surface $S$ but without self-propulsion. By applying a suitably chosen external force $\vec{F}_\mathrm{ext}$ and torque $\vec{M}_\mathrm{ext}$ one can drive the passive particle so that it moves with exactly the same translational and rotational velocities as the self-propelled particle. This is possible because $\vec{F}_\mathrm{ext}$ and $\vec{M}_\mathrm{ext}$ together have six independent components, which is sufficient for controlling the six degrees of freedom of the rigid particle. Thus, the motion of a force-free and torque-free self-propelled particle is equivalent to the motion of a passive particle with the same shape that is driven by an appropriate effective force $\vec{F}=\vec{F}_\mathrm{ext}$ and torque $\vec{M}=\vec{M}_\mathrm{ext}$. For run-and-tumble particles \cite{TailleurC2008,CatesT2013,Paoluzzi2013} the orientations of the effective force and torque are constant in the particle frame, whereas for active Brownian particles \cite{WittkowskiL2012,CatesT2013,StenhammarMAC2014} also their magnitudes are constant in the particle frame.

Due to the linearity of Stokes flow, the generalized effective force and torque vector $\vec{K}=\left(\vec{F},\vec{M}\right)$ can be written in the form $\eta \HH \vec{V}$ with the generalized velocity vector $\vec{V}=\left(\vec{v}, \vec{\omega}\right)$ and a
$6\!\times\!6$-dimensional grand resistance matrix $\HH$, which depends on the shape of the surface $S$. Thus, the validity of the concept of effective forces and torques is not affected by the specific type of self-propulsion of active particles. However, in order to relate these effective quantities directly to the physical process responsible for the self-propulsion, the details of the respective propulsion mechanism have to be known. The detailed calculations in sections \ref{mechanisch} and \ref{diffusiophoretisch} illustrate how the concept can be explicitly applied to specific situations.

\section{\label{limitations}Modifications and limitations of the concept 
of effective forces and torques}
In the following, we address the questions when the concept of effective 
forces and torques has to be modified in order to still provide a valid 
theoretical model and when it finally reaches its limitations.

While we focused on particles with constant propulsion in this 
article, the concept can also be transferred to single active Brownian 
particles with time-dependent self-propulsion \cite{Babel2014} in an 
unbounded fluid. This is particularly relevant in the context of 
run-and-tumble particles \cite{TailleurC2008,Polin2009} or when the 
swimming stroke itself results in variations of the propulsion speed.

The concept is also very useful if self-propelled particles in 
additional external fields such as gravity \cite{tenHagenKWTLB2014} are 
considered.
All external body forces that do not affect the self-propulsion itself can 
easily be included in the generalized force vector $\vec{K}$. However, if 
the intrinsic propulsion mechanism of an active particle is disturbed by 
the external field, the concept of effective forces and torques 
reaches its limitations. As an example, we refer to bimetallic nanorods 
driven by electrophoresis \cite{Paxton2004,Paxton2006} in an external 
electric field.

The situation becomes more complicated when the solvent flow field 
which is generated by the self-propelled particles
governs the particle dynamics and  has to be taken into account. The 
far-field behaviour of this solvent flow field is different from that for
a passive particle exposed to a body force. While the latter corresponds 
to a force monopole, the former is a force dipole such that
pusher- and puller-like swimmers can be distinguished. While the details of 
the solvent velocity field do not play any role for a single particle
in an unbounded fluid, they affect the motion of a particle near system 
boundaries \cite{Kreuter2013,TakagiSM,Chilukuri2014}
and at high concentrations. In particular hydrodynamic interactions 
between
different swimmers will depend on these solvent flow fields 
\cite{Kapral2008,Gompper2009,Alexander2009,Reigh2012}. Therefore the concept of 
effective forces
and torques cannot straightforwardly be applied in these situations. How exactly the effective forces and torques would change near walls or other particles is a matter of future research.

\section{\label{Schluss}Conclusions}
We have shown that effective forces and torques can be used to model the self-propulsion of microswimmers and that this concept is an appropriate and consistent theoretical framework to describe the dynamics of anisotropic active particles and to understand related experimental results. We have provided general arguments as well as specific examples for the concept of effective forces and torques. In particular, we have presented a fine-grained derivation for an L-shaped mechanical microswimmer and an explicit hydrodynamic derivation for a biaxial diffusiophoretic microswimmer. Although the detailed processes responsible for the self-propulsion of some artificial microswimmers are not fully understood yet 
\cite{Brown2014}, our general theoretical approach constitutes a powerful tool to describe the dynamics of self-propelled particles of arbitrary shape and turns out to be in good agreement with experimental observations \cite{Kuemmel:13,tenHagenKWTLB2014}.

\ack
This work was supported by the Deutsche Forschungsgemeinschaft (DFG) through the priority programme SPP 1726 on microswimmers
under contracts BE 1788/13-1 and LO 418/17-1, by the Marie Curie-Initial Training Network Comploids funded by the European Union Seventh Framework Program (FP7), by the ERC Advanced Grant INTERCOCOS (Grant No.\ 267499), and by EPSRC (Grant No.\ EP/J007404). R.W. gratefully acknowledges financial support through a Postdoctoral Research Fellowship (WI 4170/1-2) from the DFG.

\end{document}